\documentclass[10pt,conference]{IEEEtran}
\usepackage{color}
\usepackage{graphicx}
\usepackage{amsmath}
\usepackage{amsthm}
\usepackage{mathtools}
\usepackage{bm}
\usepackage{hyperref}
\usepackage{multirow}

\newcommand{\bra}[1]{{\left\langle #1 \right|}}
\newcommand{\ket}[1]{{\left| #1 \right\rangle}}
\usepackage{tikz} 
\usetikzlibrary{matrix,positioning}
\usepackage{amsfonts}
\usepackage{amssymb}

\hypersetup{
   colorlinks = true,
    urlcolor=blue
}
\begin{document}

\markboth{Anthony Wilkie}
{An angle rounding parameter initialization technique for ma-QAOA}

\title{An angle rounding parameter initialization technique for ma-QAOA}

 \author{\IEEEauthorblockN{Anthony Wilkie\IEEEauthorrefmark{1}\IEEEauthorrefmark{2}, James Ostrowski\IEEEauthorrefmark{1}, Rebekah Herrman\IEEEauthorrefmark{1}\IEEEauthorrefmark{3}
 }
 \IEEEauthorblockA{\IEEEauthorrefmark{1}Industrial and Systems Engineering, University of Tennessee at Knoxville, Knoxville, TN, USA}\thanks{
 A.W., J.O., and R.H. acknowledge the NSF award CCF-2210063.}


 Email: \IEEEauthorrefmark{2}awilkie1@vols.utk.edu, 
 \IEEEauthorrefmark{3}rherrma2@utk.edu}





\maketitle

\begin{abstract}
The multi-angle quantum approximate optimization algorithm (ma-QAOA) is a recently introduced algorithm that gives at least the same approximation ratio as the quantum approximate optimization algorithm (QAOA) and, in most cases, gives a significantly higher approximation ratio than QAOA.
One drawback to ma-QAOA is that it uses significantly more classical parameters than QAOA, so the classical optimization component more complex.
In this paper, we motivate a new parameter initialization strategy in which angles are initially randomly set to multiples of $\pi/8$ between $-\pi$ and $\pi$ and this vector is used to seed one round of BFGS.
We find that this parameter initialization strategy gives average approximation ratios of $0.900$, $0.982$, and $0.997$ for $p = 1, 2, 3$ layers of ma-QAOA.
This is comparable to the average approximation ratios of ma-QAOA where the optimal parameters are found using BFGS with 1 random starting seed, which are $0.900$, $0.982$, and $0.996$.
We also test another parameter initialization strategy in which angles corresponding to maximal degree vertices in the graph are set to 0 while all other are randomly initialized to random multiples of $\pi/8$.
Using this strategy, the average approximation ratios are $0.897$, $0.984$, and $0.997$.
\end{abstract}



\section{Introduction}
The quantum approximate optimization algorithm (QAOA) is a hybrid quantum-classical algorithm that uses parameterized circuits with a classical optimization subroutine to solve combinatorial optimization problems \cite{farhi2014quantum}.
The algorithm evolves an initial state $\ket{s}$ under alternating cost ($U(C, \gamma) = e^{-i C \gamma}$) and mixing ($U(B, \beta) = e^{-i B \beta}$) unitaries for parameters (also called angles) $\beta, \gamma \in [-\pi, \pi)$ classically chosen to maximize the expected value of $U(C, \gamma)$. 
The QAOA output state is 
\begin{equation*}
    \ket{\gamma, \beta} = U(B, \beta_p)U(C, \gamma_p) \ldots U(B,\beta_1)U(C, \gamma_1)\ket{s}
\end{equation*}
\noindent where $C$ is problem dependent, and in some cases, $B$ and $\ket{s}$ are as well.

The MaxCut problem is one of the most widely studied CO problems in the context of QAOA \cite{guerreschi2019qaoa, herrman2021impact, crooks2018performance, wang2018quantum}, however the algorithm can be applied to any CO problem with an Ising formulation \cite{Ozaeta_2022}. 
The MaxCut problem is a graph problem in which the vertices of a graph $G$ are partitioned into two sets such that the number of edges with endpoints in each set is maximized. 
In typical QAOA studies on the MaxCut problem,

\begin{equation*}
    C = \frac{1}{2} \sum_{ij \in E(G)}(I - Z_iZ_j),
\end{equation*}
where $Z_i$ is the Pauli-z operator acting on qubit $i$.
The choice of mixer $B$ can vary, but it is typically the sum of Pauli-x matrices acting on each qubit. 
The metric of success for the algorithm is the approximation ratio, which is the ratio of $\bra{\gamma, \beta}C\ket{\gamma, \beta} = \langle C \rangle$ over the size of an optimal problem solution.
The approximation ratio is a measure of how well QAOA does at finding the optimal value $C^*$ of the problem, and is calculated by
\begin{equation*}
    AR = \frac{\langle C \rangle}{C^*}.
\end{equation*}

QAOA is guaranteed to converge in the limit as $p \rightarrow \infty$ when the initial state $\ket{s}$ is the lowest energy eigenstate of $B$ \cite{farhi2014quantum}, however since the algorithm is currently implemented on NISQ devices \cite{Zhou_2020QAOAImplementationPerformance, earnest2021pulse}, minimizing the number of iterations it takes to converge on an optimal solution is critical. 
In fact, recent work suggests that QAOA cannot beat the classical Goemans-Williamson algorithm on 3-regular graphs in fewer than 11 iterations \cite{wurtz2021fixed}. 
Thus, variants of QAOA such as ST-QAOA, X-QAOA, ma-QAOA, and sym-QAOA have been introduced in an effort to decrease circuit depth \cite{herrman2021multi, Shi2022, vijendran2023expressive, wurtz2021classically}. 
This work will focus on ma-QAOA, which is a QAOA variant in which each term of $C$ and $B$ are given their own variational parameter, so $U(C, \vec{\gamma}) = e^{-i C_k \gamma_k}$ and $U(B, \vec{\beta}) = e^{-i B_j \beta_j}$. 
The expected value for $p$ iterations of ma-QAOA will be denoted $\langle C \rangle_p^{\mathrm{ma}}$ throughout this work, and the approximation ratio will be denoted $AR$.
Since ma-QAOA requires significantly more variational parameters, the classical optimization subroutine becomes more challenging, especially when $p>1$  \cite{gaidai2023performance}. 
However, in their work, the authors of \cite{herrman2021multi} show that a significant proportion of the angles that were found to optimize ma-QAOA on all connected 8-vertex graphs were $0$, indicating that clauses can be dropped from $U(C, \gamma)$ and $U(B, \beta)$.

Several parameter initialization strategies for QAOA have been introduced \cite{Sack2021, Sack2023, Sud2022, sureshbabu2023parameter}, however ma-QAOA parameter initialization strategies have not been well-studied\cite{gaidai2023performance}. 
In this work, we develop an ma-QAOA initial angle setting heuristic to attempt to alleviate the challenges associated with finding optimal ma-QAOA parameters while still maintaining a comparable approximation ratio. 
In Section~\ref{sec:angledata}, we first analyze patterns in optimal ma-QAOA angles.
These inspire an angle rounding heuristic described in Sec.~\ref{sec:heuristic} and an ma-QAOA angle initialization strategy, which is detailed in Section~\ref{sec:angleinitialization}.
In Section~\ref{sec:max_degree}, we test a third angle initialization strategy inspired by noticing that many of the optimal angles from the other strategies are 0 when they correspond to maximum degree vertices in the graph.
We close with future research directions in Section~\ref{sec:discussion}.

\section{Angle data and graph structure}\label{sec:angledata}
In this work, we analyze the angles that optimize ma-QAOA on all connected, non-isomorphic four-vertex and eight-vertex graphs (over 11,000 graphs), as well as 50 randomly created non-isomorphic graphs of 9, 10, 11, and 12 vertices, respectively.
All data was generated using the code found at \cite{wilkiecode}, which is a modification of the code found at \cite{maqaoarepo}, and then angles outside of the interval $(-\pi, \pi)$ were normalized to be within the interval.
The parameter optimization subroutine for the four-vertex graphs uses the Broyden-Fletcher-Goldfarb-Shanno algorithm \cite{NumericalRecipesBFGS} with 1 random initial seed. 

\begin{figure*}
    \includegraphics[width=1.0\textwidth]{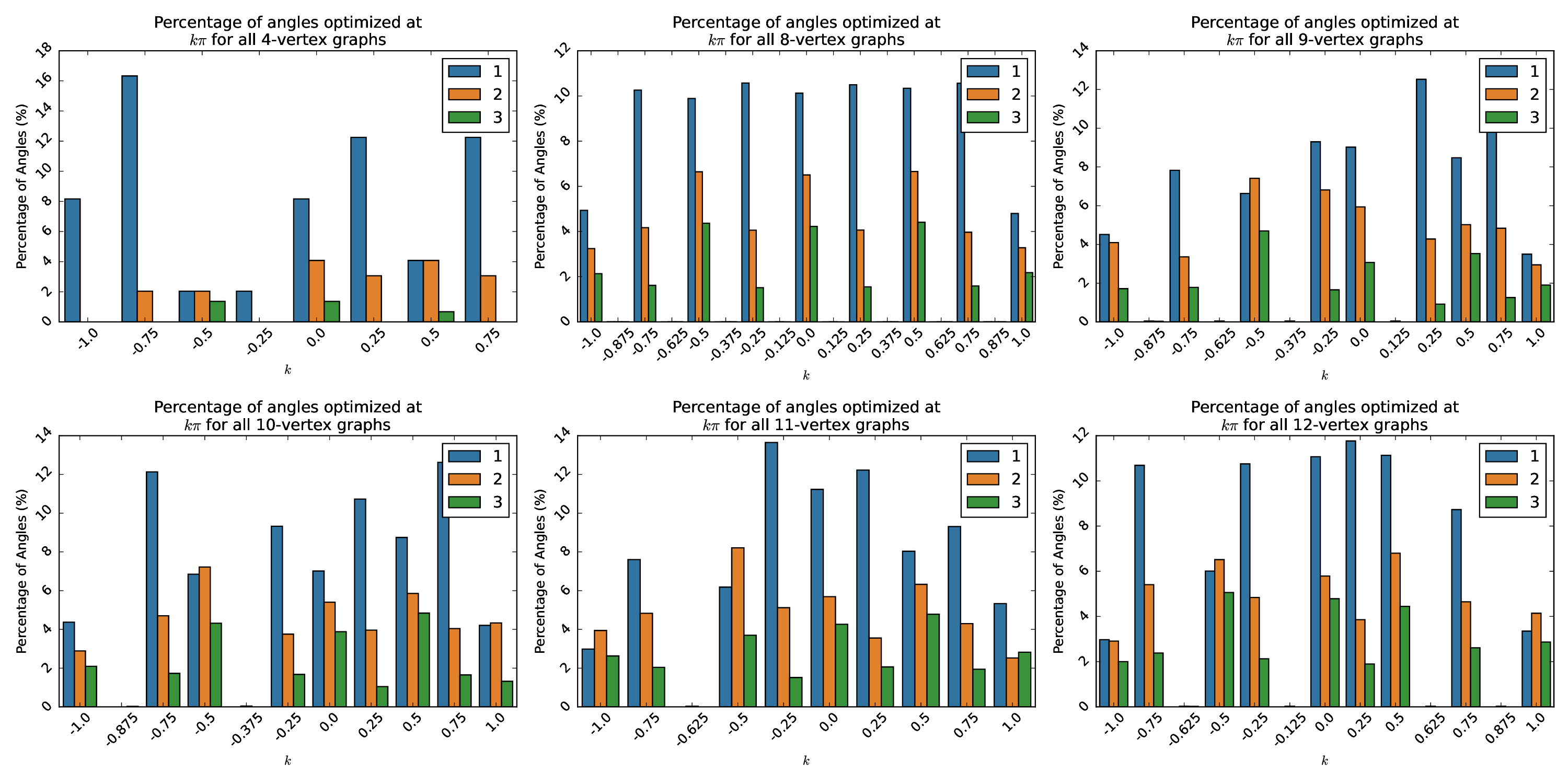}
   \caption{The percent of $\beta_v$ and $\gamma_{i,j}$ that are optimized at $k\pi$ for all graphs tested and up to three layers of ma-QAOA.}
    \label{fig:percent_angles_rounded}
\end{figure*}

The four-vertex graph dataset has forty-nine angles, twenty-four of which are $\beta$ (four $\beta$ angles for each of the six graphs).
The eight-vertex dataset has $249,156$ angles, $88,936$ of which are $\beta$ angles and the rest are $\gamma$ angles.
Interestingly, in both datasets, the $\gamma$ and $\beta$ that maximize $\langle C \rangle_1^{\mathrm{ma}}$ tend to concentrate around multiples of $\pi/8$.
Around 65.31\% of the $\gamma$ and $\beta$ that maximized the expected value for the four-vertex graph dataset were multiples of $\pi/8$ while approximately 81.99\% of the eight-vertex angles were multiples of $\pi/8$ (within three decimal places) with $p=1$ layer of ma-QAOA.
The distribution of angles in the eight-vertex case is relatively symmetric about $0$, as well, which can be seen in Fig.~\ref{fig:percent_angles_rounded}.


 \section{An angle rounding heuristic}\label{sec:heuristic}
 Approximately 65.31\% (four-vertex graphs) and 81.99\% (eight-vertex graphs) of the $\beta_i$ and $\gamma_{uv}$ that maximize $AR$ for $p=1$, as found in \cite{wilkiecode}, are multiples of $\pi/8$.
Even though a significant percentage of angles that maximize $AR$ are not multiples of $\pi/8$, it is possible that there exist angles that are multiples of $\pi/8$ that still yield the same expected value.
Thus, we round all angles that are not multiples of $\pi/8$ to the nearest multiple of $\pi/8$ and calculate the new approximation ratio.
We call the approximation ratio calculated using angles that were rounded $AR^{\mathrm{rounded\;init}}_{\mathrm{no\;opt}}$, with $\mathrm{no\;opt}$ indicating the angles were not optimized and plugged directly back into the ma-QAOA circuit.

\begin{figure*}
    \includegraphics[width=1.0\textwidth]{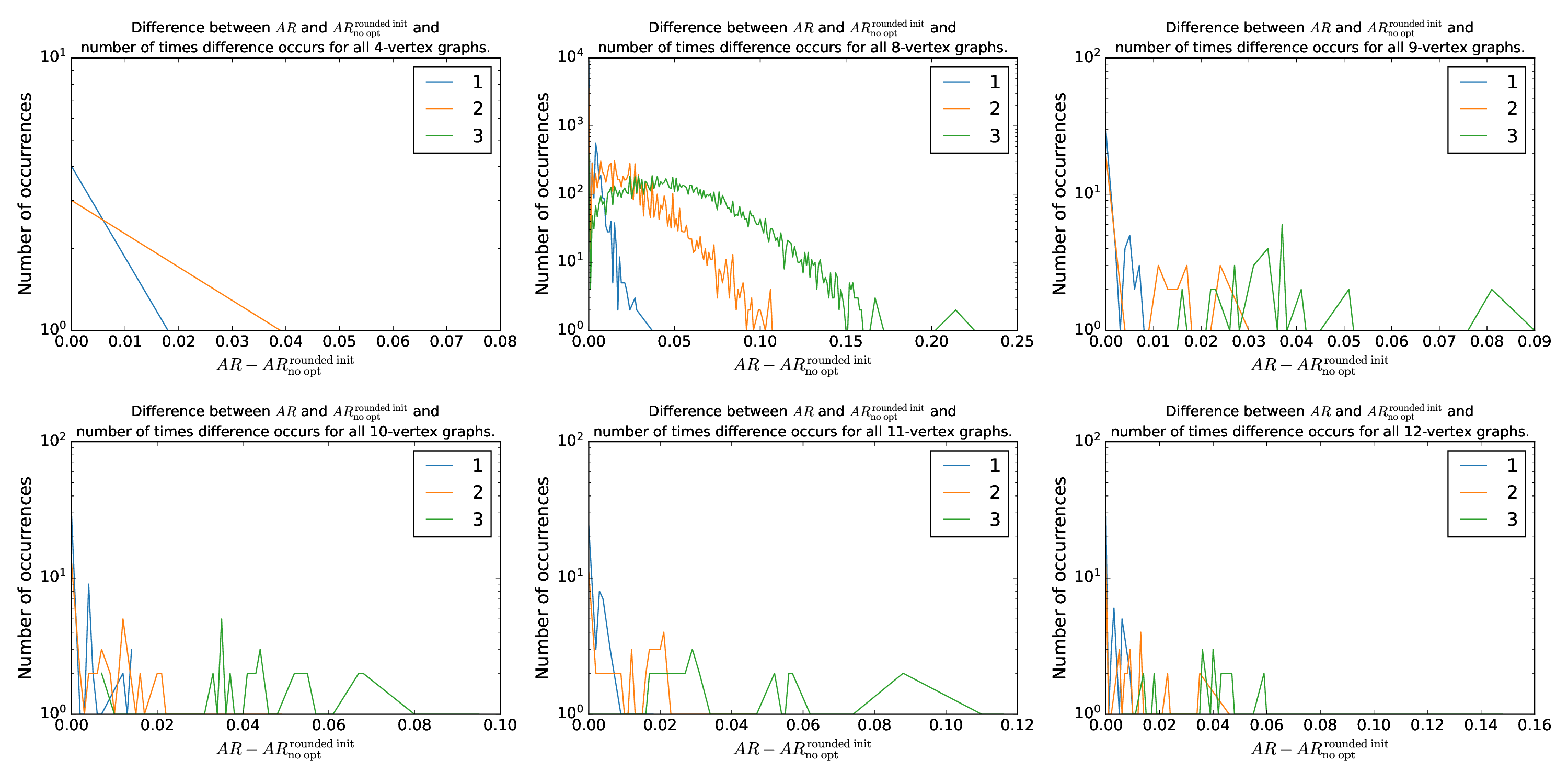}
   \caption{Difference between  $AR$ and $AR^{\mathrm{rounded\;init}}_{\mathrm{no\;opt}}$ and the number of times the difference occurs for all graphs tested for up to three layers of ma-QAOA.}
    \label{fig:evroundeddiff}
\end{figure*}

Approximately 80.668\%, 31.458\%, and 3.471\% of all the graphs tested have the same $AR$ (rounded to 3 decimal places) when the angles are rounded as they had when the angles are not rounded, for $p = 1$, $2$, and $3$ respectively.
The data in Fig.~\ref{fig:evroundeddiff} shows the difference per graph between the original approximation ratios $AR$ and the ones where the rounded angles we plugged back into the circuit $AR^{\mathrm{rounded\;init}}_{\mathrm{no\;opt}}$.
In general, $AR^{\mathrm{rounded\;init}}_{\mathrm{no\;opt}}$ does not perform as well as $AR$, with the difference between them increasing as $p$ increases.
This rounding technique does assume we have the optimal angles a priori, which may not always be the case.

\section{An angle rounding parameter initialization strategy}\label{sec:angleinitialization}
In this section, we present a heuristic for initializing ma-QAOA angles based on the data in Section~\ref{sec:angledata}.
From Fig.~\ref{fig:percent_angles_rounded}, we see that a significant proportion of the angles that maximize $AR$ are multiples of $\pi/8$.
We propose a heuristic that initializes all angles to random multiples of $\pi/8$ in the interval $[-\pi, \pi]$ and use this starting point to seed one round of BFGS for parameter optimization.

\begin{figure*}
  \includegraphics[width=1.0\textwidth]{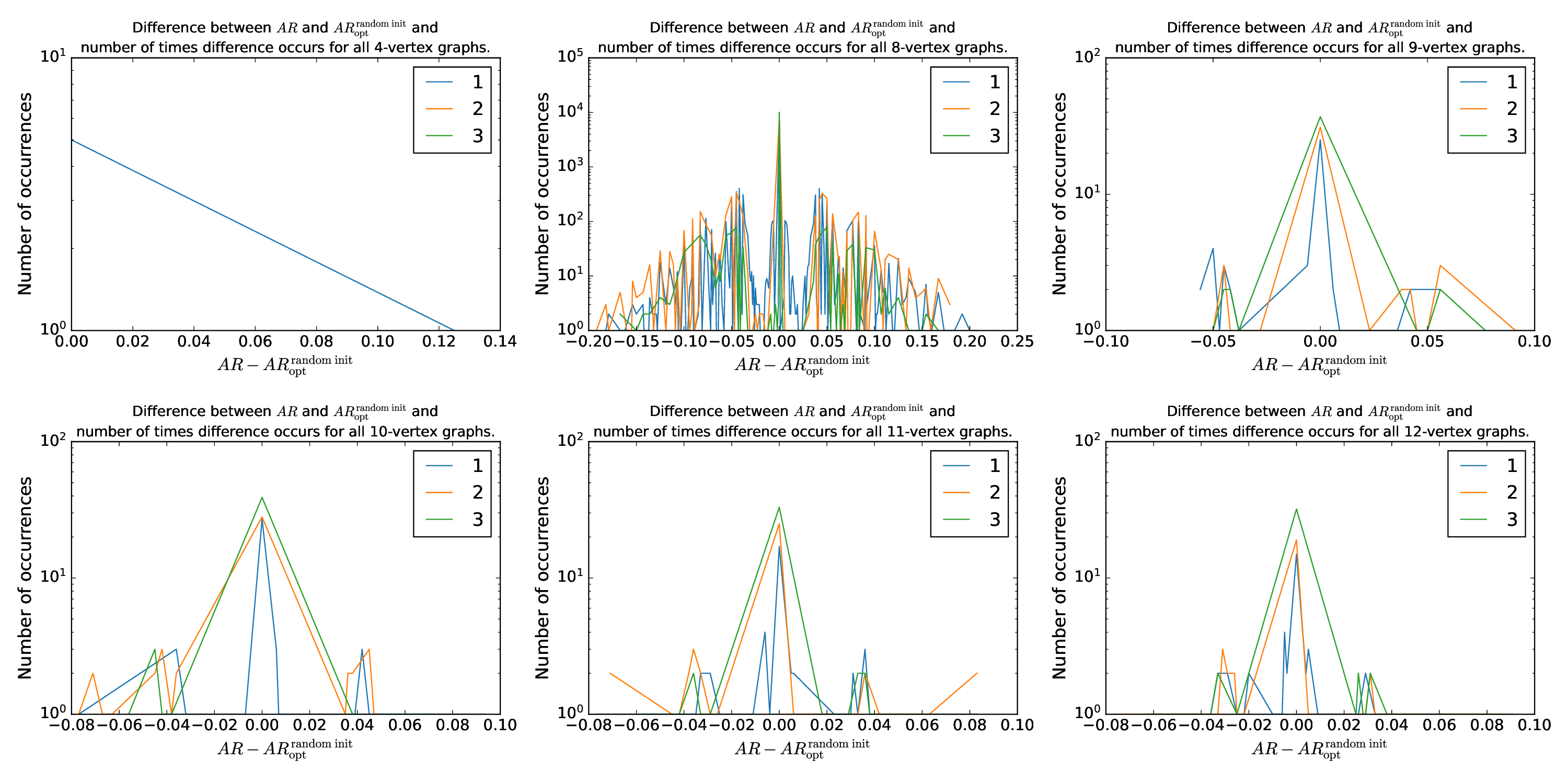}
  \caption{Difference between $AR$ and $AR^{\mathrm{random\;init}}_{\mathrm{opt}}$ for all graphs tested and up to three layers of ma-QAOA.}
  \label{fig:diff_ar_eight_vertex_random}
\end{figure*}

When using this initialization strategy, we find that 75.828\%, 83.088\%, and 95.160\% of all the graphs tested with this angle heuristic had approximation ratios greater than or equal to those without it, for $p = 1, 2, 3$ respectively.
In Table~\ref{tab:mean_ar_table}, we can see that the mean approximation ratios are also higher using $p=1$ in the case of the $9$, $11$, and $12$ vertex graphs.
In Fig.~\ref{fig:diff_ar_eight_vertex_random}, we see that the difference between $AR$ and $AR^{\mathrm{random\;init}}_{\mathrm{opt}}$ (the approximation ratio using the parameter initialization technique) tends to peak at $0$ for all graphs, except in the case of the $4$-vertex graphs.
Thus, using one round of BFGS to optimize parameters that are initially seeded to be random multiples of $\pi/8$ performs well and can requires significantly less classical optimization effort than using no initialization strategies and just using one randomly seeded round of BFGS to find the optimal parameters.


\section{Angles Corresponding to Max Degree Vertices}\label{sec:max_degree}
\begin{figure*}
  \includegraphics[width=1\textwidth]{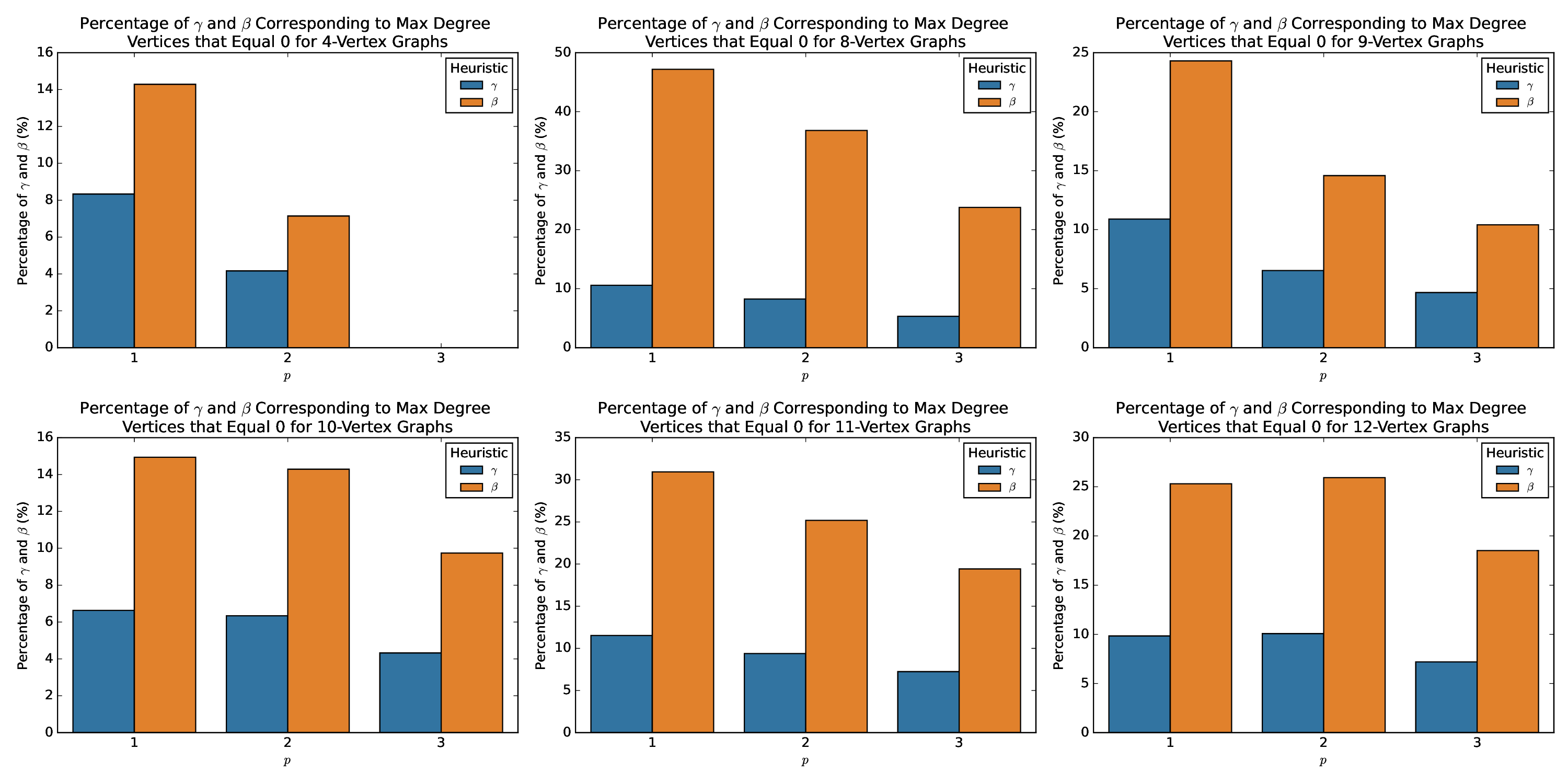}
  \caption{Percentage of $\gamma$ and $\beta$ angles corresponding to max degree vertices that equal 0.}
  \label{fig:percent_gamma_beta_max_degree}
\end{figure*}

\begin{figure*}
  \includegraphics[width=1.0\textwidth]{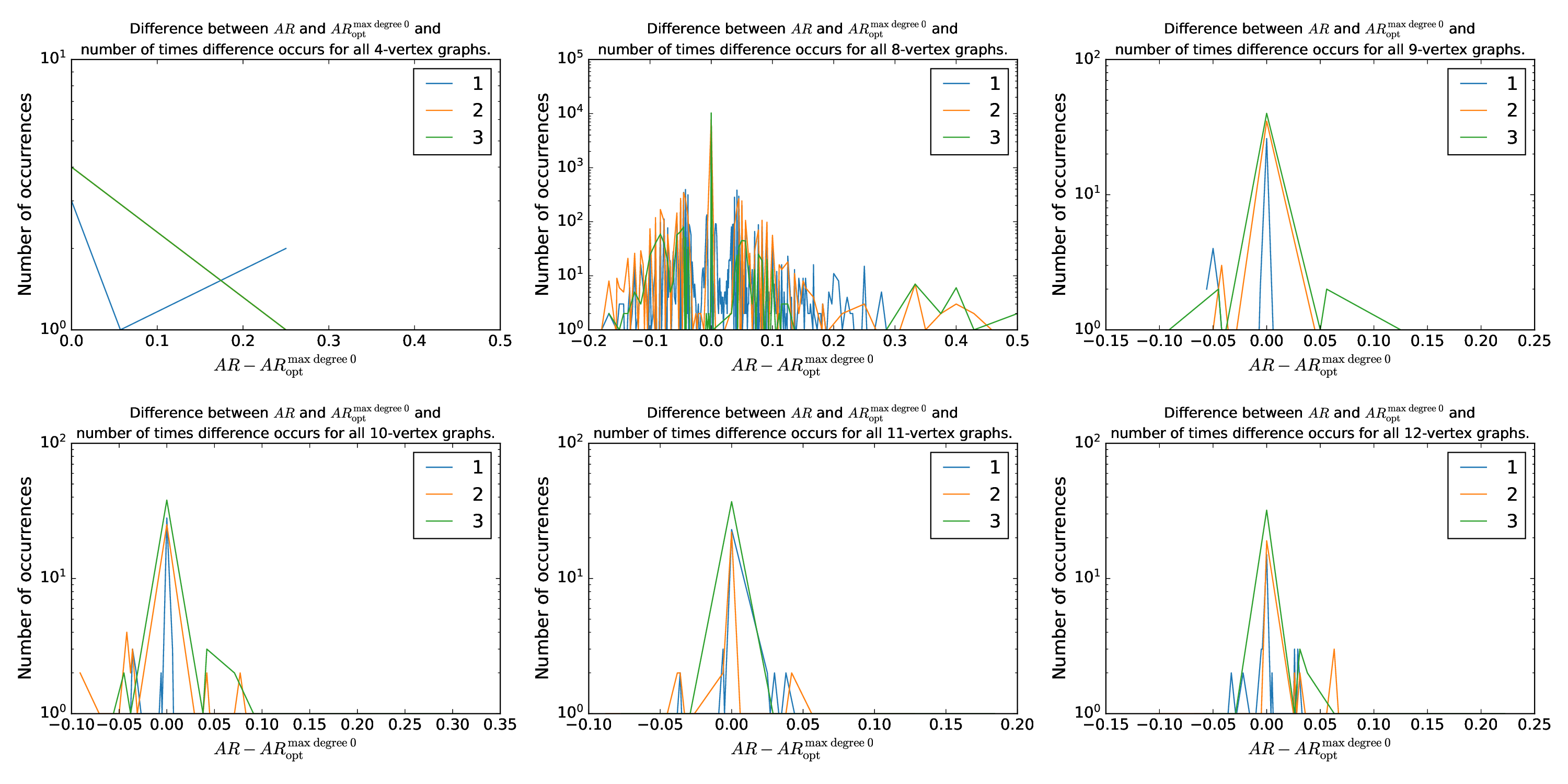}
  \caption{Difference between the approximation ratios for normal ma-QAOA and the approximation ratios for the max degree set to 0 angle heuristic with one step of optimization for up to three layers of ma-QAOA.}
  \label{fig:diff_ar_eight_vertex_max_degree_0}
\end{figure*}

In the search of finding better angle initialization techniques, a relationship between the $\gamma$ and $\beta$ angles that are equal to 0 and correspond to maximum degree vertices in a graph was found, shown in Fig.~\ref{fig:percent_gamma_beta_max_degree}.
In these figures, for each of the different angle heuristics tested, there is a significant portion of the angles that have this property, with the number of angles decreasing as $p$ increases.
The importance of this relationship is that for a maximum degree vertex $v$ in a graph, there are $\mathrm{deg}(v)$ many $\gamma$'s and one $\beta$ that need to be optimized.
If these angles are equal to $0$, then the gates these angles parameterize become identity gates, and hence can be discarded from the circuit, reducing circuit depth.
In Fig.~\ref{fig:diff_ar_eight_vertex_max_degree_0}, we test the heuristic of setting these angles that correspond to maximum degree vertices to 0 and randomize the other angles to multiples of $\pi/8$.
The case where the angles are not optimized and plugged directly into the circuit does, just as with the rounded and random initializations, does not perform well, with differences in approximation ratios $AR$ and $AR^{\mathrm{max\;degree\;0}}_{\mathrm{no\;opt}}$ averaging at 0.257, 0.340, and 0.354 for $p = 1, 2, 3$, respectively.
However, when the angles are optimized, the results improve and in some cases performing better that normal ma-QAOA, with the difference $AR - AR^{\mathrm{max\;degree\;0}}_{\mathrm{opt}}$ averaging at $0.003$, $-0.002$, and $0$, for $p=1, 2, 3$, respectively.
The mean and standard deviation of the approximation ratios for both the un-optimized and optimized cases can be found in Table~\ref{tab:mean_ar_table}.



\begin{table*}
\caption{The mean approximation ratios for each angle initialization heuristic tested for each number of nodes in the graph and up to three layers of ma-QAOA.}
\label{tab:mean_ar_table}
\centering
\begin{tabular}{|c|*{8}{c|}}
\hline
Vertices & $p$ & Normal & Rounded No Opt & Random Opt & Max Degree 0 Opt \\
\hline
\hline
\multirow{3}{2em}{4} & 1 & $0.931 \pm 0.111$ & $0.923 \pm 0.114$ & $0.910 \pm 0.107$ & $0.838 \pm 0.193$ \\
 & 2 & $1.000 \pm 0.000$ & $0.973 \pm 0.031$ & $1.000 \pm 0.000$ & $0.875 \pm 0.209$ \\
 & 3 & $1.000 \pm 0.000$ & $0.954 \pm 0.024$ & $1.000 \pm 0.000$ & $0.875 \pm 0.209$ \\
\hline
\multirow{3}{2em}{8} & 1 & $0.900 \pm 0.060$ & $0.899 \pm 0.060$ & $0.900 \pm 0.060$ & $0.898 \pm 0.064$ \\
 & 2 & $0.983 \pm 0.032$ & $0.966 \pm 0.036$ & $0.983 \pm 0.032$ & $0.985 \pm 0.034$ \\
 & 3 & $0.997 \pm 0.015$ & $0.947 \pm 0.034$ & $0.997 \pm 0.015$ & $0.997 \pm 0.021$ \\
\hline
\multirow{3}{2em}{9} & 1 & $0.881 \pm 0.049$ & $0.879 \pm 0.050$ & $0.888 \pm 0.052$ & $0.879 \pm 0.065$ \\
 & 2 & $0.970 \pm 0.036$ & $0.955 \pm 0.038$ & $0.963 \pm 0.037$ & $0.977 \pm 0.034$ \\
 & 3 & $0.987 \pm 0.027$ & $0.948 \pm 0.030$ & $0.991 \pm 0.023$ & $0.989 \pm 0.027$ \\
\hline
\multirow{3}{2em}{10} & 1 & $0.875 \pm 0.052$ & $0.872 \pm 0.054$ & $0.872 \pm 0.053$ & $0.875 \pm 0.050$ \\
 & 2 & $0.958 \pm 0.031$ & $0.946 \pm 0.034$ & $0.964 \pm 0.033$ & $0.959 \pm 0.039$ \\
 & 3 & $0.992 \pm 0.020$ & $0.948 \pm 0.027$ & $0.991 \pm 0.023$ & $0.981 \pm 0.048$ \\
\hline
\multirow{3}{2em}{11} & 1 & $0.874 \pm 0.040$ & $0.872 \pm 0.041$ & $0.878 \pm 0.035$ & $0.875 \pm 0.036$ \\
 & 2 & $0.962 \pm 0.037$ & $0.949 \pm 0.037$ & $0.962 \pm 0.027$ & $0.957 \pm 0.032$ \\
 & 3 & $0.986 \pm 0.022$ & $0.946 \pm 0.033$ & $0.985 \pm 0.022$ & $0.979 \pm 0.042$ \\
\hline
\multirow{3}{2em}{12} & 1 & $0.863 \pm 0.046$ & $0.861 \pm 0.047$ & $0.866 \pm 0.044$ & $0.861 \pm 0.045$ \\
 & 2 & $0.956 \pm 0.036$ & $0.944 \pm 0.038$ & $0.949 \pm 0.034$ & $0.952 \pm 0.035$ \\
 & 3 & $0.984 \pm 0.020$ & $0.942 \pm 0.034$ & $0.981 \pm 0.025$ & $0.965 \pm 0.056$ \\
\hline
\end{tabular}

\end{table*}

\section{Discussion}\label{sec:discussion}
In this work, we present an angle rounding parameter initialization scheme for ma-QAOA where ma-QAOA initial angles are randomly selected to be multiples of $\pi/8$ between $-\pi$ and $\pi$, and then BFGS is seeded from this vector and allowed to search the parameter space for optimal solutions.
The technique requires few initial seeds for BFGS to attain an average approximation ratio of $0.900$, $0.982$, and $0.997$ for $p = 1, 2, 3$ layers of ma-QAOA.
This is within 0\% of the average approximation ratio of ma-QAOA with no parameter initialization technique for all 3 $p$ tested, which is $0.900$, $0.982$, and $0.996$ on the same dataset. 
We also tested a parameter initialization technique determined by setting angles corresponding to maximal degree vertices equal to 0, which yielded the average approximation ratios $0.897$, $0.984$, and $0.997$.
These results show that proper parameter initialization can perform well without the need to perform multiple runs of the classical optimizer to achieve good approximation ratios.

Interestingly, the subgraphs induced by unique optimal edge angles tend to be forests.
This is reminiscent of ST-QAOA, where a spanning tree of the underlying graph is found and given its own QAOA parameters.
This work suggests that instead of using spanning trees, setting distinct angles for forests may provide higher approximation ratios.
Future work includes determining how to leverage forests in parameter selection or initialization routines.
Other future work includes leveraging problem symmetry for faster training, such as in \cite{sauvage2024building, shaydulin2021exploiting}.

\section*{Acknowledgements}
The authors would like to acknowledge NSF award CCF-2210063. The authors would like to thank Phillip Lotshaw for his insight.

\renewcommand\refname{References Cited}
\bibliography{references}
\bibliographystyle{IEEEtran}

\end{document}